\documentclass[preprint,proceedings]{rmaa}

\usepackage{paralist}

\usepackage{psfrag,color}

\def\lesssim{{_ <\atop{^\sim}}}
\def\grtsim{{_ >\atop{^\sim}}}
\def\msun{\mbox{M$_\odot$}}
\def\lsun{\mbox{L$_\odot$}}
\def\kms{\mbox{kms$^{-1}$}}
\def\LCDM{\mbox{$\Lambda$CDM}}

\SetYear{2003}
\SetConfTitle{Galaxy Evolution: Theory and Observations}

\title{Physical processes behind the morphological Hubble sequence} 

 \RescaleTitleLengths{0.7}
\author{Claudio Firmani\altaffilmark{1,2} and Vladimir Avila-Reese\altaffilmark{2}}

\altaffiltext{1}{Osservatorio Astronomico di Brera, Italy.}
\altaffiltext{2}{Instituto de Astronom\'\i{}a, UNAM, M\'exico.}
\shortauthor{Firmani \& Avila-Reese}
\shorttitle{Physics beyond the Hubble Sequence}

\fulladdresses{
\item Claudio Firmani: Osservatorio Astronomico di Brera, via E. Bianchi 46, 
I-23807 Merate, Italy

\item Vladimir Avila-Reese: Instituto de Astronom\'{i}a, U.N.A.M, A.P. 70-264, 
04510, M\'{e}xico, D.F., M\'{e}xico }

\listofauthors{C. Firmani \& V. Avila-Reese}
\indexauthor{Firmani, C.}
\indexauthor{Avila-Reese, V.}

\resumen{Presentamos una rese\~na de la formaci\'on y evoluci\'on de las 
galaxias haciendo \'enfasis en los factores f\'{\i}sicos relevantes para 
entender el origen de la secuencia de Hubble. Nos concentramos en las
predicciones del escenario jer\'arquico de Materia Obscura Fr\'{\i}a (MOF) 
y su confrontaci\'on con las observaciones. Describimos los 
procesos de agregaci\'on  de masa de los halos de MOF, los procesos en la 
materia bari\'onica atrapada por los halos, as\'{\i} como la formaci\'on y 
evoluci\'on de los discos y esferoides. Los aciertos y las
debilidades de la teor\'{\i}a son analizados. La evoluci\'on de los discos 
parece implicar un proceso cont\'{\i}nuo y extendido inducido por las 
condiciones cosmol\'ogicas iniciales, mientras que 
los esferoides derivan probablemente de eventos violentos en los cuales 
compiten varios procesos astrof\'{\i}sicos.
}

\abstract{The study of formation and evolution of galaxies is reviewed,
with emphasis on the physical factors which are important to understand
the origin of the Hubble sequence. We concentrate on predictions of 
the hierarchical Cold Dark Matter (CDM) scenario
and their confrontation with observations. The mass assembling of the CDM 
halos, the baryonic processes within them, and the evolution of disks and 
spheroids are described. The successes and shortcomings are discussed. 
Disk evolution seems to be a quiescent and extended process driven by the 
cosmological initial conditions, while spheroids are formed probably in 
violent events, where several astrophysical processes are competing.}

\addkeyword{galaxies: ellipticals}
\addkeyword{galaxies: evolution}
\addkeyword{galaxies: spirals}
\addkeyword{cosmology: dark matter}

\begin{document}
\maketitle

\section{Introduction}
\label{sec:intro}

The Cold Dark Matter (CDM) cosmological model, inspired by the inflationary 
theory,  has triggered the development of a provocative {\it deductive} 
scenario of galaxy formation and evolution. According to this scenario, galaxies
form within non-dissipative dark matter halos, which assemble hierarchically
from primordial density fluctuations (White \& Rees 1978). The deductive 
approach offers a valuable theoretical frame for interpreting the galaxy 
phenomena and motivates new confrontations between theory and observations. 
In parallel, from the observed properties of Milky Way (MW) and nearby galaxies, 
chemical and spectrophotometric evolutionary models allowed us to infer, by
an {\it inductive} approach, fundamental aspects of galaxy evolution. 
More recently, the observation of galaxies at different redshifts is making 
possible to trace directly the luminous, structural and dynamical evolution 
of whole populations of galaxies. 

The question of how the Hubble morphological sequence originated and 
evolved in its complex relation with the environment is one of the most exciting 
challenges of extragalactic astronomy. The Hubble sequence (HS) of galaxies is 
based on a morphological classification, but several physical properties also 
change along this sequence (Roberts \& Haynes 1994), a relevant morphological 
indicator being the bulge-to-disk ratio (Hubble 1936; Simien \& de Vaucouleurs 
1986). Therefore, it is natural to concentrate our attention on the physical 
nature of disks and spheroids, and on how galaxy structures do emerge in the 
context of a unified cosmological scenario.

While the evolution of disks appears gradual and extended in time, 
spheroids (elliptical galaxies and bulges) seem to concentrate their star 
formation activity upon a violent, early phase of their evolution.
According to hierarchical galaxy formation schemes, 
most of the stars originated in disks, while subsequent processes of 
dynamical nature transformed some of these disks into spheroids. Nowadays,
the increasing information of the high-redshift universe is opening new 
perspectives for the understanding of the formation of spheroids, its
connection with ultra luminous infrared galaxies (ULIGs), quasars (QSOs), 
active galactic nuclei (AGNs), as well as for the interplay of 
galaxy formation with the intergalactic medium.

In this review our attention will be focused on the formation and evolution
of disks and spheroids in the hierarchical CDM scenario. Some recent reviews
on observational and evolution aspects of the HS can be found in 
Roberts \& Haynes (1994), Ellis (1999, 2001), Freeman (1999), van den 
Bergh (2002).  Previous reviews on the physical origin of the HS can be 
found in Silk \& Wyse (1993) and Pfenniger (1996).

The plan of this review is first to discuss the origin and properties of the 
galactic dark matter halos, as well as the gas infall and galaxy formation 
processes within them (\S\S 2 and 3). Then, the status of models of disk 
and spheroid formation in the hierarchical scenario and their comparison 
with observations will be presented in \S\S 4 and 5, respectively. In \S 6 
a sketch of the unified phenomenon of galaxy formation in the cosmological 
context is presented, and in \S 7 the concluding remarks as well as some 
perspectives for the future are enumerated.

\section{Dark matter processes}

In a CDM universe, the formation of cosmic structures is 
governed by gravitational processes. At large scales
the influence of baryons is negligible. Starting from a Gaussian 
density fluctuation field with a given processed power spectrum, 
the gravitational clustering of the fluctuations is followed to 
the present epoch by means of cosmological N-body simulations,
and the results are confronted with observations (for recent numerical 
results see e.g., Jenkins et al. 1998; Evrard et al. 2002).
The completion of large surveys of galaxies and clusters of galaxies,
the measurements of sub-degree anisotropies in the microwave 
background radiation, and the detection of cosmological SNe
improved significantly our understanding of the large-scale structure 
and the mass-energy composition of the Universe (for a recent review see Guzzo 
2002 and the references therein). The agreement between observations and 
the CDM predictions is remarkable (see Frenk, this volume). 
The ``concordance'' \LCDM\ cosmological model (Bahcall et al. 1999)
emerges as the favorite one. For this model, the universe is flat with
the following approximate values of the cosmological parameters:   
$\Omega_{\rm bar}=0.04$, $\Omega_{\rm CDM}=0.26$, $\Omega_{\Lambda}=0.7$, 
$\sigma_8 = 0.9$, and $H_0=65 \kms$ Mpc$^{-1}$. In this review, with the 
generic term CDM we will refer to the \LCDM\ model.

At the scales of galaxies and clusters of galaxies, where high resolution 
is required in the simulations, an extensive work on formation, mass function, 
structure, and evolution of the CDM matter halos has been done in the last 
decade. Analytical and semi-analytical approaches, but mainly 
numerical N-body simulations, were used. The dark halos are the backbone of 
the galaxy formation models. Following, we discuss some results which 
appear relevant for the properties of galaxies:

{\it i)} The shape of the {\it mass function} of CDM halos is approximately 
similar to that of the observed Schechter luminosity function of galaxies
(e.g., Press \& Schechter 1974; Lacey \& Cole 1993). The semi-analytical 
models show that the main problem is at the faint end of the luminosity 
function (e.g., Kauffmann et al. 1993; Cole et al. 2000; Somerville \& Primack 
1999); however, reionization and feedback may possibly solve the conflict 
(Benson et al. 2002a, see also this volume). 

{\it ii)} The average {\it density profiles} of CDM halos are described 
typically by a universal two parameter profile, both depending ultimately 
only on the halo mass (Navarro, Frenk, \& White 1997, hereafter NFW). 
Less massive halos tend to be more concentrated than the more massive ones. 
However, for a given mass, the halo density profiles show a scatter around
the NFW profile. This scatter correlates with the halo mass aggregation 
history (MAH), in the sense that halos assembled earlier are more 
concentrated (Avila-Reese et al. 1998,1999; Wechsler et al. 2002).
Some dependence on the environment has also been reported (but see 
Lemson \& Kauffmann 1999). The CDM halos are too concentrated and their 
inner density profiles are cuspy, in apparent disagreement with 
observations, mainly the inner rotation curves of dwarf and LSB galaxies 
(Moore 1994; Burkert 1995; see also Bosma, de Blok, and Col\'{\i}n et al. 
in this volume).  At galaxy-cluster scales, the inferred halo inner density 
profiles, under the uncertainties, typically are fitted by both the NFW and 
the pseudo-isothermal profiles. 
Observations seem to show that, from dwarf to galaxy-cluster scales, the 
central halo density is poorly dependent on mass, and the core radius increases 
roughly proportional to the maximum circular velocity V$_{\rm max}$ 
(Firmani et al. 2000, 2001). Figure 1 presents the approximate range
of values of the halo central density and core radius vs. V$_{\rm max}$
inferred from observations. 
Another potential problem of the CDM halos is that the number of subhalos 
within MW-sized halos overwhelms the number of observed satellite 
galaxies by a large factor (Kauffmann et al. 1993; Klypin et al. 1999; Moore 
et al. 1999). Besides, the large population of satellite subhalos could have
a dramatic effect on the dynamics of the galaxy disks ( Moore et al. 1999;
Colpi, Mayer \& Governato 1999).
Owing to the success of the CDM model at large scales, only minor
modifications to the model has been proposed in order to solve these 
potential problems. For example, in a Warm Dark Matter (WDM) scenario 
with particle masses of $\sim 0.6-1$ KeV, the satellite velocity function in 
MW-sized halos is well reproduced, while at larger scales the predictions
are similar to CDM (Col\'{\i}n et al 2000). However, the inner 
density profiles of galaxy-sized and larger halos are similar to their CDM 
counterparts; even the small subhalos show density profiles well 
fitted by the NFW profile, although with concentrations lower than
predicted by CDM (Avila-Reese et al. 2001; Bode, Ostriker \& Turok 2001). 
Conversely,
self-interacting dark matter (SIDM) with a velocity-dependent 
cross sections, $\sigma_{DM}=0.5-1.0 \ (100 \kms$/V$_{\rm max})$ cm$^2$/gr, 
produces inner halo density profiles in agreement with observations
at all scales (Fig. 1), but the substructure remains similar to CDM (Firmani
et al. 2001; Col\'{\i}n et al. 2002, see also this volume). 
The substructure problem may be actually alleviated by the reionization,
which certainly has to be taken into account in the formation of dwarf galaxies
(Bullock et al. 2000; Benson et al. 2002a).

iii) The {\it angular momentum} distribution in most of CDM halos 
seems to be well parametrized by a universal function, and the disks formed 
within them, assuming detailed angular momentum conservation, are roughly 
exponential (Bullock et al. 2001b). The global spin parameter $\lambda$ has a 
lognormal distribution and is approximately independent of the cosmology, mass, 
and environment (e.g. Catelan \& Theuns 1996). 
Two mechanisms for the origin of the halo angular momentum seem to compete: 
linear tidal torques and orbital angular momentum transfer of merging satellites 
(e.g., Peebles 1969; Vitvistkaya et al. 2002; Maller, Dekel \& Somerville 2002). 
The role of angular momentum in formation of galaxies is crucial. 
More work should be done, for example, on the symmetry  and degree of 
alignment of the angular momentum distribution in the halos 
(see van den Bosch et al. 2001 for recent results) and on the angular momentum 
dependence on environment.

\begin{figure}[!t]
\includegraphics[width=1.05\columnwidth]{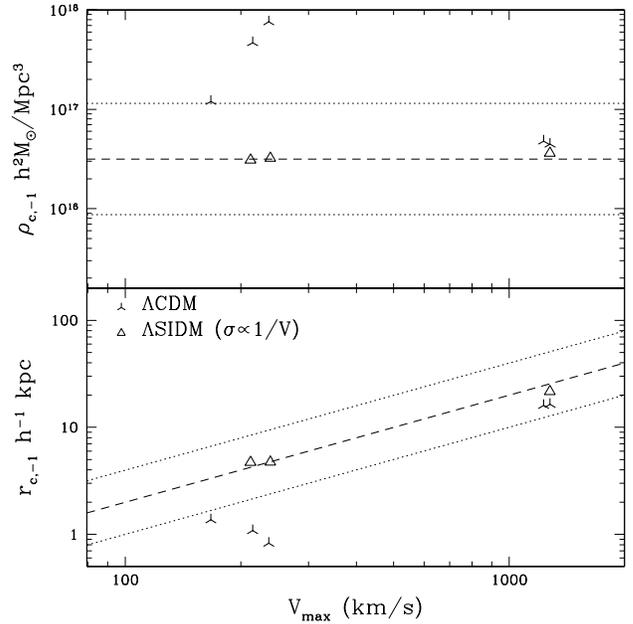}
\caption{Central density and core radius in this plot are defined
as the density and radius where the halo profile slope becomes
steeper than -1. The dotted lines encompass roughly the 
2 $\sigma$ dispersion of the observational inferences of 
$\rho_{c,-1}$ and $r_{c,-1}$ vs. V$_{max}$ for dwarf and LSB
galaxies and several clusters of galaxies (see Col\'{\i}n et al. 2002
for the source references). Squeletal and open triangles are high-resolution 
halo simulations for the $\Lambda$CDM and $\Lambda$SIDM models;
the latter one is for a cross section inversely proportional to 
the relative velocity.}
\end{figure}
 
iv) The hierarchical {\it mass aggregation history} (MAH) of CDM halos 
scatters around its average, because they emerge from a stochastic density field. 
On average, the less massive halos tend to assemble a given fraction of their 
mass at earlier epochs than the more massive halos (Avila-Reese et al. 1998). 
The halos may grow by a variety of merging regimes going from smooth
mass accretion to violent major mergers (e.g., Salvador-Sol\'e et al. 1998).
This variety of merging regimes is relevant to the morphology of the galaxies
formed within the halos. Cosmological simulations show that (i) most of the 
$z=0$ galaxy-sized halos are not contained within larger halos nor have
close massive companions, and that (ii) most of the mass of these halos has
been aggregated by smooth accretion (Avila-Reese et al. 1999; 
Somerville \& Kolatt 2000; Gottl\"ober et al. 2001). 
Nevertheless, the pair abundance and the major merger rate increase with 
$z$, and this increase is  faster for cluster halos than for field halos 
(Gottl\"ober et al. 2001).  The predicted major merging rates agree with 
those inferred from accurate statistics of galaxy pairs. From the fraction 
of normal galaxies in close companions (with separations less than 
50 kpch$^{-1}$) inferred from observations
at $z=0$ and $z=0.3$ (Patton et al. 2000, 2002), and assuming an average 
merging time of $\sim 1$ Gyr, we estimate that the major merging rate at the 
present epoch is $\sim 0.01$ Gyr$^{-1}$ for halos in the range of 
$0.1-2.0 \ 10^{12}\msun$, while at $z=0.3$ the rate increased to 
$\sim 0.018$ Gyr$^{-1}$. These values are only slightly lower than predictions 
for the $\Lambda$CDM model, suggesting that the model can be used to predict 
the increasing merging rate at earlier epochs.
The fact that at high redshifts the major merging rate is
much higher than at the present epoch, may be at the basis of 
early spheroid formation, as well as energetic phenomena like QSOs and
submillimeter sources (see \S\S 5 and 6).

Using cosmological N-body simulations, we have learnt much about the 
clustering and properties of collapsed CDM structures. Nevertheless, 
important details about the innermost, less resolved regions of dark 
halos, where just luminous galaxies form, as well as about the angular 
momentum, the MAHs 
and their dependences on the environment, remain still unexplored. 
On the other hand, the potential conflicts of CDM at small scales are 
still open. More observational and theoretical studies are necessary 
to understand the depth of these conflicts.

\section{Baryon matter processes}

Cosmological simulations including gas show that at $z=0$ only 
$30-40\%$ of the universal baryon matter is within collapsed halos; the 
rest is in the form of warm-hot IGM in the filaments and voids (Dav\'e
et al. 2001). Recent observations tend to confirm this prediction 
(e.g., Manucci, this volume). Not all the gas trapped within the galactic halos 
ends up in the central galaxy: for large halos (M$\grtsim 10^{12}\msun$) the 
cooling time may be longer than the Hubble time, while for small halos 
(M$\lesssim 10^{10}\msun$) a large fraction of gas may be expelled due to 
feedback. Owing to the combination of cooling and feedback, galaxies incorporate 
on average only half of the baryons originally trapped within halos 
(van den Bosch 2002).
Therefore, for a universal fraction $\Omega_{\rm bar}/\Omega_{\rm CDM}\approx 
0.15$, the actual fraction of matter trapped in galaxies, f$_{\rm gal}$, is only 
$\approx 0.15\times 0.4\times 0.5\ = 0.03$.

The physics of baryons within the collapsing and merging dark matter
halos is a highly complex process. In some cases, instead of radiative 
cooling, {\it turbulent dissipation in a multi-phase regime} 
(unexplored in detail yet) dominates. The process is further complicated
by the gravitational fragmentation and consequent transformation of gas
into stars. The energy and momentum input on gas from stars introduces a 
feedback. The feedback establishes a control and self-regulates the SF. 
The self-regulation may be either {\bf (a)} at the level of the disk ISM, 
where, according to the nature of the feedback, a variety of regimes appear, 
ranging from stationary SF to bursting SF (e.g., Firmani \& Tutukov 1994), or
{\bf (b)} at the level of the whole intrahalo medium, giving rise to a huge
hot gas halo around the luminous galaxy (e.g., White \& Frenk 1991;
Benson et al. 2000). Models of galaxy evolution in the cosmological context 
used either case (a) for disks (Firmani et al. 1996, Firmani \& Avila-Reese 
2000; Avila-Reese \& Firmani 2000; van den Bosch 2000) or case (b) 
(e.g., Kauffmann et al. 1993, 1999; Cole et al. 1994, 2000; Baugh et al. 1996). 

Since dark halos have some angular momentum, during the smooth mass accretion 
phase, the dissipating gas falls until it reaches centrifugal equilibrium: disks 
are a generic prediction of the hierarchical scenario (Kauffmann et al. 1993). 
As the disk ISM is dense, highly dissipative and with shielding magnetic 
fields, the energy injection on gas is confined to a few disk scale heights  
(Avila-Reese \& V\'azquez-Semadeni 2001, and references therein). 
The lack of bright X-ray halos around spirals (Benson et al. 2000) reinforces 
the idea that case (a) rather than case (b) applies for disk galaxies. 
On the other hand, case (b) probably is working in spheroids where the low 
internal gas density makes dissipation inefficient. 
It seems also that cooling is more efficient than that calculated by simple 
approaches in the semi-analytic models (Toft et al. 2002). 
When the halo angular momentum is sufficiently low or the gas angular momentum 
loss is efficient, then highly concentrated disks may form. 
These disks are gravitationaly unstable and a spheroid may be produced by a 
secular process, which involves the formation and dissolution of a bar
(Combes et al 1990; Norman et al. 1996; Merrifield \& Kuijken 1999; Valenzuela
\& Klypin 2003).

As mentioned in \S 2, dark halos grow by a variety of merging regimes. 
Major mergers induce collisions between the central luminous galaxies.
In this case, SF follows a bursting regime (Firmani \& Tutukov 1994), feedback is
very efficient, and the morphological Hubble type is strongly affected. 
Collisions between galaxies are indeed able to produce dynamically hot spheroids 
(Toomre \& Toomre 1972; Schweizer 1982; Barnes 1988; Hernquist 1990, 1992). 
If the colliding galaxies are gas rich disks, then gas dissipation 
makes it possible to concentrate the spheroid up to the observed densities 
(Mihos and Hernquist 1994,1996; see also Hibbard and Yun 1999). 
Collisions between galaxies and secular formation from bar dissolution 
represent two competitive processes for the formation of spheroids.

Using the semi-analytical models, several authors extended the hierarchical 
merging of CDM halos to a merging picture of luminous disk/bulge/elliptical 
galaxy formation (e.g., Baugh et al. 1996; Cole et al. 2000). According to 
this picture, the driver of galaxy evolution, as far as concerns the HS, 
is the dark halo merging. Cosmological simulations indeed show the possibility 
of rapid galaxy morphological transformations due to intermittent changes of 
quiescent phases of disk accretion with violent phases of spheroid formation 
by major mergers (e.g., Steinmetz \& Navarro 2002). However, cosmological
simulations also show that most galaxies are formed actually by smooth
accretion rather than mergers (Murali et al. 2000). This is in agreement
with the predominance in number of disk galaxies in the local universe. 

All these physical ingredients are involved in galaxy evolution, and are 
crucial to the understanding of the origin of the HS. It is now natural to 
concentrate our attention on the physics of disks and spheroids.

\RescaleSecLengths{0.6} 

\section{Disks}

A model aimed to study the physics and evolution of disk galaxies should 
include a self-consistent and local description of dark matter processes, and
gas dissipation, infall, SF, stellar evolution, and feedback.
Under the assumptions of (i) dominance of accretion rather than mergers, 
and (ii) detailed angular momentum conservation during gas infall, 
several authors have been able to
reproduce the main properties and correlations of disk galaxies in 
the hierarchical CDM scenario. The former assumption is motivated  by the 
fact that disks are dynamically fragile objects and collisions with other 
galaxies heat and thicken the stellar disk beyond the level observed
at present (T\'oth \& Ostriker 1992). The latter assumption is argueable,
and is currently among the key questions of disk formation models 
(see e.g., van den Bosch et al. 2002). Following, we discuss the main 
predictions of models of disk formation inside CDM halos. A seminal
paper on this subject was that of Fall  \& Efstathiou (1980).

\subsection{Disk models at $z=0$}

The {\it stellar surface density profile} is roughly exponential, but with 
an excess at the inner and outer regions (Dalcanton et al. 1997; Firmani \& 
Avila-Reese 2000; Bullock et al. 2001a; van den Bosch 2002). The inner cusp  
may be crucial for the secular bulge formation. The surface density of the 
disks depends mainly on the spin parameter $\lambda$: the angular momentum
drives the sequence from high surface brightness (HSB) to low surface brightness 
(LSB) galaxies. The surface density also depends on f$_{gal}$. Mainly due to 
the inside-out formation, the disks have {\it negative color index and 
metallicity gradients} in agreement with observations (Avila Reese \& Firmani 
2000; Boissier \& Prantzos 1999, 2001).

The {\it rotation curves} are roughly flat. The shape of rotation curves 
depends on the disk surface density (Mo, Mao \& White 1998; Firmani \& 
Avila-Reese 2000). 
LSB galaxy models are dominated by the dark halo and even for the extreme HSB 
models, the CDM halo is competitive with the disk at the maximum rotation 
velocity. For the inner region of normal HSB galaxy rotation curves to 
be dominated by disk (maximum disk case), a soft core in the CDM halo is 
necessary (Avila-Reese et al. 2002; see also Salucci \& Burkert 2000; 
Salucci 2001). 

The slope of the infrared {\it Tully-Fisher relation (TFR)} is interpreted 
as a consequence of the mass-velocity relation of the CDM halos, 
determined in turn by the power spectrum of fluctuations 
(e.g., Mo et al. 1998; Avila-Reese et al. 1998, 1999; Steinmetz \& Navarro 
1999; Firmani \& Avila-Reese 2000;  Buchalter et al. 2001).
The slope of the halo mass-velocity relation derived from theory is 
$\approx3.2-3.4$, which agrees 
with the TFR slope in the infrared bands found by most observers 
(e.g., Gavazzi 1993; Giovanelli et al. 1997; Tully et al. 1998; 
Tully \& Pierce 2000). After the formation of the disk and the stars
within it, the zero point of the model TFR is slightly fainter than 
observations for the concordance $\Lambda$CDM cosmology (see Fig. 2), 
and is far too faint for the SCDM one. 
The introduction of soft halo cores improves the comparison with observations 
(Firmani \& Avila-Reese 2000; Mo \& Mao 2001).
The N-body+hydrodynamical simulations predict a too faint zero point even 
for the $\Lambda$CDM model (e.g., Steinmetz \& Navarro 2000), but this 
is mainly because the disks in these simulations are formed highly concentrated 
(angular momentum catastrophe), implying very peaked rotation curves.

Variations in f$_{gal}$ and $\lambda$ do not affect significantly the TFR.
For a larger f$_{gal}$, the luminosity will be higher, but V$_{\rm max}$ 
also results larger. A smaller $\lambda$ implies a more dominant disk
component and therefore a larger V$_{\rm max}$ for a given mass; however, 
the SF efficiency is higher in such a way that the stellar
mass (luminosity) for a fixed baryonic mass will be also larger (Firmani 
\& Avila-Reese 2000). This is the way we interpret {\it why the TFRs of 
LSB and HSB galaxies are similar and why the residuals of the TFR do not 
correlate with the residuals of the magnitude-radius relation} 
(Courteau \& Rix 1999). 
We also find that the main source of intrinsic scatter in the TFR derives 
from the scatter in the halo concentrations due to the stochastic nature 
of the MAHs (Eiseinstein \& Loeb 1996); the average scatter we measure is 
$\sim 0.3$ mag, close to the observational estimates. 

If bulges are modeled as the gravitationally unstable stellar disk region (secular
scenario), then higher surface brightness galaxies have larger bulge-to-disk 
ratios (van den Bosch 1998; Avila-Reese \& Firmani 2000). The models
reproduce the {\it main correlations along the disk HS}: the higher 
the SB and the redder the disk, the larger the bulge-to-disk ratio and the 
lower the disk gas fraction. Bulges formed secularly seem to be in 
agreement with observations, at least for late-type galaxies (see for
a review Wyse et al. 1997).

Models can be tested even in the {\it solar neighborhood}. For a MW 
fiducial model (based on the statistically most probable cosmological MAH), 
the predicted evolution of the SF rate in the solar neighborhood agrees within
the uncertainty with the evolution derived from the HR diagram provided by the 
Hipparco satellite (Hern\'andez, Avila-Reese \& Firmani 2001). 
On the other hand, chemical and photometric models, starting from the observed 
distributions in the MW, infer infall, structural and SF conditions similar 
to those expected from the hierarchical scenario (e.g., Boissier \& Prantzos 
1999; Prantzos, this volume). However, the inner dynamics again reveals a 
conflict: 
the CDM halo concentrates too much mass in the central regions in such a way that 
the shapes of the model and observed MW rotation curves do not agree. 
A soft core of size as inferred from observations of LSB and dwarf galaxies, 
eliminates the discrepancy.

\subsection{Disk evolution}

The sizes of disks in the hierarchical scenario decrease significantly 
with redshift (Mao, Mo et al. 1998; Giallongo et al. 1999; Avila-Reese 
\& Firmani 2001). 
Interpretation of data regarding disk size and SB evolution is controversial. 
Roche et al. (1998) using $HST$ data concluded that since $z\approx 1$, spirals 
suffered both size and luminosity evolution, while Simard et al. (1999) concluded 
that data show no size evolution. Bouwens \& Silk (2002) show that the SB 
distribution of disk galaxies evolves strongly. They re-analyzed the data of 
Simard et al. introducing new corrections for SB selection bias and found 
size evolution in the data. The question is open.

The SF history of disk models is driven by both the gas accretion rate 
determined by the MAH, and the disk surface density determined by $\lambda$ 
(Avila-Reese \& Firmani 2001). Merging could also play a relevant role 
(Kauffmann et al. 2001). The increase of the SF rate (and $B-$band luminosity) 
with $z$ inferred from observations of normal spirals (e.g., Lilly et al. 1997; 
Abraham et al. 1999) is slightly steeper than our model predictions. The 
integral colors of the models become bluer toward the past, in agreement with 
observations.

\begin{figure}[!t]
\includegraphics[width=\columnwidth]{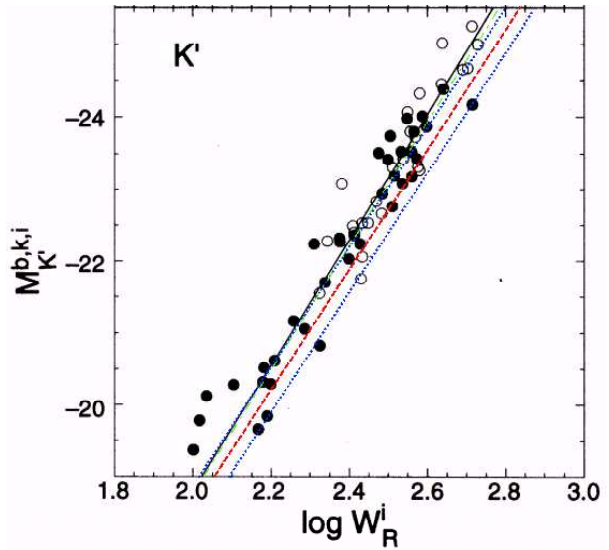}
\caption{Tully-Fisher relation in the band $K$ as inferred
by Tully et al. (1998). Solid line is the linear regression to 
the data. Dashed line is the TFR from models of galaxy evolution 
in the $\Lambda$CDM cosmogony (Firmani \& Avila-Reese 2000); 
detailed angular momentum conservation was assumed. Dotted lines 
correspond to the model scatter ($\pm 1-\sigma$ dispersion). The
dot-dashed lines is the TFR for the same models, but introducing a soft
core in the halos to be in accord with the rotation curve inner shapes
of LSB and dwarf galaxies. The agreement with observations is excellent. 
The TFR is an imprint of the density fluctuation power spectum.}
\end{figure}

The evolution of the TFR is a highly debated topic and apparently it depends
on whether the studied sample is dominated by small or normal spirals. 
Model predictions show that the slope of the $B-$band TFR decreases and 
the zero point becomes brighter at higher $z$'s (Fig. 3; see also 
Avila-Reese \& Firmani 2001).
Comparing with the observed sample of Vogt et al. (1997) at $<z>=0.54$,
the agreement in the zero point is reasonable (for the $\Lambda$CDM cosmology). 
On the other hand models show that the zero-point of TFR in the near 
infrared band evolves in an opposite way: at $z=1$ it is fainter than at $z=0$.
(see Fig. 3) 
This can be understood taking into account that galaxy evolution traces halo 
evolution: the halo mass (scaling with the disk mass traced by the infrared 
luminosity) decreases significantly from $z=0$ to $z=1$, while V$_{\rm max}$ 
decreases only moderately. In the case of the $B-$band, since the accretion 
rate peaks at $z\sim 1-2$, the star formation rate, and therefore L$_B$, are 
(slightly) higher at these reshifts. The evolution of the comoving number 
density of halos is connected with the TFR; a deficit of bright (massive) 
galaxies at larger redshifts translates into a dimming of the zero-point of 
the infrared TFR (Bullock et al. 2001a). The evolution of the TFR up to high 
redshifts ($z\sim 3$) can be a potentially powerful discriminator of galaxy 
formation models (Buchalter, Jim\'enez \& Kamionkowski 2001).

\begin{figure}[!t]
\includegraphics[width=\columnwidth]{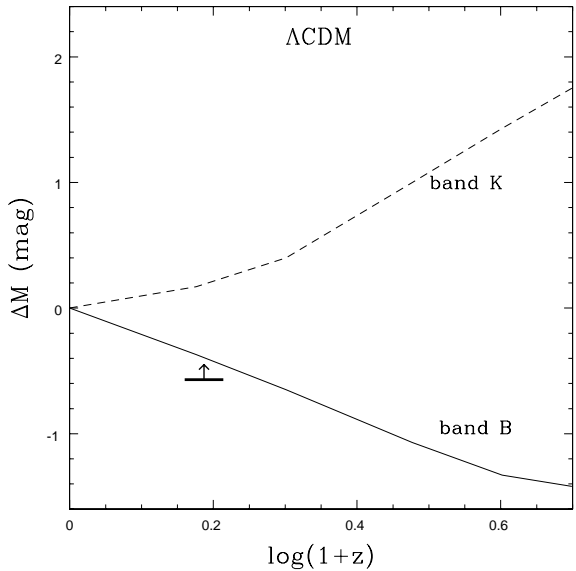}
\caption{Evolution of the model TFR in the bands $B$ and $K$. The 
slopes at different redshifts were fixed to the corresponing ones at
$z=0$. The zero-point becomes brighter at higher redshifts in 
the band $B$, while in the band $K$ the behaviour is opposite (see
text). The arrow indicates the lower limit inferred from observations
by Vogt et al. (1997); we have passed their data to the $\Lambda$CDM
cosmology used here.}
\end{figure}

\subsection{Drivers of the disk Hubble sequence}

The main properties, correlations and evolution features of disk 
galaxies formed within CDM halos, under the assumption mentioned 
at the beginning of this section, arise from the combination of three
cosmological factors and their statistical distributions: 
the {\it virial mass}, the {\it MAH}, and the {\it spin parameter 
$\lambda$}. The former determines the scale. The latter ones determine 
the SF history and the intensive properties, and are at the basis of 
the disk HS correlations, suggesting a biparametrical nature for this 
sequence (Avila-Reese \& Firmani 2000). Observations tend to confirm 
this biparametrical nature (e.g, de Blok \& McGaugh 1997). 
The MAH drives the gas infall rate, which determines mainly the galaxy 
color indexes. The $\lambda$ parameter determines mainly the 
disk SB, and strongly influences the rotation curve shape and the 
bulge-to-disk ratio (secular bulge). 
The mass galaxy fraction, f$_{\rm gal}$, which influences some disk 
properties, is an astrophysical parameter determined mainly by the halo 
gas dissipation and feedback efficiencies; it may vary from galaxy to 
galaxy, but its value on average is around 0.03 (see \S 3). 
The shapes of the rotation curves and the SB distribution are better 
reproduced jointly just for this value (Avila-Reese et al. 2002). 

An important ingredient, which begins now to be included in 
the CDM models of disk evolution, is the dynamics related to
bars, spiral arms and minor mergers (e.g., see the contributions
by Valenzuela, Athanassoula, Curir, Bertschik and others in this 
volume). This ingredient, highly connected to the main disk properties 
and processes described above, is at the basis of the ``showy''
disk morphological signatures, which characterize the tuning
fork of the disk HS, and may play some role in the morphological
transformation due to secular processes.

\section{Spheroids}

In spite of their simple shape, spheroids reveal a very complex nature. 
Some memory of their formation lies in their dynamical structure and 
stellar population. The spheroid complexity is strengthened by the 
systematical presence of super massive black holes (SMBHs) at their nuclear 
regions, with rather well defined correlations between the SMBH mass and 
the spheroid luminosity or velocity dispersion (Magorrian et al. 1998; 
Ferrarese \& Merrit 2000; Gebhardt et al. 2000), and the bulge concentration
(Graham et al. 2001; Graham, this volume). 
AGN and QSO activity seem to be intimately associated with SMBHs. 
On the other hand, observations reveal that the ultra high
luminosity infrared galaxies (ULIGs) show evidence of recent galaxy 
collisions (Sanders and Mirabel 1996); therefore ULIGs may also 
be linked to the formation of spheroids. 

\subsection{Relevant observational properties and correlations}

Radius, velocity dispersion and surface brightness are related by the fundamental 
plane (FP) (Bender et al. 1992), which is basically related to the virial 
theorem and to the M/L ratio inside the effective radius. The use of NIR 
photometry (Pahre 1998, Pahre et al. 1998a) reduces the effects of the 
SF history on the M/L ratio and traces the gravitational contribution 
of dark matter. A homology breaking (dynamical [Busarello et al. 1997; Graham
\& Colless 1997] and/or 
photometric [Bertin et al. 2002]), connected to the spheroid formation and 
evolution, introduces effects difficult to evaluate. 
The observed evolution of the FP intercept up to $z=0.4$ suggests a passive 
evolution for the stellar population, with a formation redshift of $1.0-1.5$ 
(Pahre et al. 1998b, Treu et al. 2001). 
The change in slope with $z$ is not clear, but the possibility that massive
ellipticals could be younger than the less massive ones seems to be ruled out. 
The scatter of the FP does not show any correlation with metallicity, redshift 
(age effect), or environment (field or clusters).

The brightness of giant spheroids decreases with size (Kormendy relation, KR;
Kormendy 1977; Pahre et al. 1998a), implying that the mean density of spheroids 
decreases with their mass. It is interesting to note that a similar tendency 
is shown by dark virialized halos in a hierarchical cosmogony. However, at 
present it is not clear if the density-mass relation of galaxy spheroids 
can be explained in terms of the dark halo properties alone or by other processes 
related to the gas hydrodynamics or to the SF. The observed luminosity evolution 
agrees with the passive evolution of a star population formed at $z\gtrsim1$. 
The Faber-Jackson relation (FJR) may be understood as a combination of the FP and 
the KR.

The $Mg_2$-velocity dispersion relation shows mainly how the metallicity 
increases with the mass. The color-magnitude relation involves both metallicity 
and age. The evolution up to $z=1$ of the zero point and of the scatter of 
this relation suggests a formation redshift $\gtrsim2$ for most of the field 
and cluster ellipticals (see Peebles 2003) (even if some uncertainty derives
from the sample selection).

From the shape of their isophotes, elliptical galaxies are divided into disky and 
boxy. It has been suggested that the origin of this dichotomy is related to the 
mass ratio of the merged disk progenitors (Naab et al. 1999).
High resolution observations of the central region of luminous, slowly rotating 
boxy ellipticals (mainly members of clusters), identify a shallow core, while in 
faint rapidly rotating disky ellipticals (mainly field galaxies), peaked 
power-law density profiles were found  (Faber et al. 1997; Carollo et al. 1997). 

\subsection{Theoretical aspects}

The unification of the physics implied in all of these properties and phenomena 
is an appealing challenge.
Gas-rich mergers (e.g. two disks), besides heating and thickening the 
preexisting disk stellar population, removes angular momentum from gas and 
allows the gas to concentrate further in the center.  The low
angular momentum gas, which dissipates probably in a turbulent regime, 
will form a compact disk, where stars are born in a high density environment, 
rich in dust and molecules. An intense burst of star formation takes place 
driven by a radiation pressure (Eddington) self-regulation, while the bulk of 
UV radiation is converted to FIR by dust (Firmani \& Tutukov 1994). 
This phase may be identified as an ULIG. In the central region gas reaches very 
high density, probably due to a gravothermal catastrophe. This condition 
favors the growth of the preexisting black hole through an Eddington
(Silk \& Rees 1998; Fabian 1999; Nulsen \& Fabian 1999; Blanford 1999, see 
also Merrifield et al. 2000) or even super-Eddington regime (Begelman 2001). 
The accretion, hidden by gas and dust, swallows only a minor fraction of the 
infalling gas. When the SMBH reaches a critical mass, the gas outflow 
becomes catastrophic and most of the gas is expelled. Probably, a small 
fraction of gas is retained in an accretion disk which continues to 
feed the SMBH during a few millions years (Burkert \& Silk 2000). The 
expulsion of gas 
ends the FIR emission phase and allows the AGN to appear in all its 
brightness. The QSO phase lasts the time sufficient to exhaust any residual 
gas feeding. After this ultra-high luminous phase finishes, the SMBH enters 
in a dormant phase in the heart of the stellar spheroid, which will appear as a 
disky elliptical galaxy with a power law inner density profile.

This scenario gives some idea about the role of merging on the formation 
of spheroids. However, other astrophysical processes may influence the further 
evolution of spheroids. Gravitation and cooling push gas to inflow. As shown 
by Ciotti et al. (1991), the gas heating due to SN Ia is enough to refund 
the energy lost by cooling and to produce supersonic galactic winds, at least 
in the early phases after the SF burst. As the SN Ia rate decreases with time, 
the wind changes to an outflow regime. Thereafter, if the spheroid is sufficiently 
massive, a cooling catastrophe in the inner region converts outflow into inflow. 
Once inflow is established, the fate of gas in the inner region is uncertain. 
Low mass SF driven by thermal instabilities can drop out gas from the flow 
(Mathews \& Brighenti 1999a, 1999b), otherwise infalling gas starts to feed 
the central SMBH, which enters into an AGN phase (Ciotti \& Ostriker 1997, 
2001). In the latter case, the gas energy balance is strongly influenced by 
the AGN energy injection. Gas inflow at some time is suddenly converted to an 
outflow, while SMBH returns to its dormant phase. At this point the cycle 
outflow-inflow-AGN repeats again. This conjecture, based on cooling flow, 
identifies an AGN driving mechanism alternative to the one based on major 
mergers. The cooling-flow driving mechanism for AGN finds some support in the 
observations of Schade et al. (2000) and Dunlop et al. (2001), who did not 
see strong evidence of major merging in their sample of AGN host spheroids. 
Several aspects of this process have to be studied taking into account, besides 
the stellar mass loss, the gas accretion predicted by the hierarchical scenario.

Several pieces of evidence support the idea that ULIGs represent the connection 
between mergers and the dust enshrouded formation of spheroids and AGNs. Even if 
ULIGs at present are rare systems and their contribution to the cosmic energy 
budget is negligible, at high redshift this contribution grows and becomes 
dominant (similar to AGNs) (Sanders and Mirabel 1996; Krishna \& Biermann 1998; 
Lilly et al. 1999; Dunlop et al. 2001; Granato et al. 2001).
Most ULIGs show evidence of strongly interacting disk galaxies. 
The FIR luminosity increases with decreasing projected nuclear separation. 
A double nucleus is present in the majority of ''cool'' systems 
$(f_{25}/f_{60}<0.2)$, while a single nucleus is more frequent in ''warm'' 
systems. The fact that ULIGs represent ellipticals in formation is supported 
by the evidence that these objects fall near the FP of disky ellipticals 
(Genzel et al. 2001).

The sequence: merger between disk galaxies $\Rightarrow$ ''cool" ULIG 
$\Rightarrow$ ''warm" ULIG $\Rightarrow$ ''IR excess" QSO $\Rightarrow$ QSO, 
is supported by the following arguments (Sanders 2001; Kim et al. 2002; 
Veilleux et al. 2002). A gradual transition from FIR to UV dominated spectral 
energy distribution (from ''cool" to ''warm") is accompanied by the appearance 
of a ''big blue bump" (characteristic of optically selected QSOs) 
and nuclear superwinds. QSOs with FIR excesses (in transition from ''warm" ULIGs) 
appear related to ''disturbed" hosts, while E-like ULIGs and AGNs seem more 
likely to lie in hosts of similar masses. Hard X-rays will provide more 
information about the SMBH growth in the ULIG nuclei (Fabian 1999; Wilman 
et al. 2000).

This scenario is consistent with a simple demographic argument. 
From {\it HST}, Lilly et al. (1996) estimated a local blue luminosity density 
of $10^7 \lsun/$Mpc$^{-3}$ Assuming $<M/L_B> \cong 14 \msun/\lsun$ 
(see Dwek et al. 1998), the luminous matter density is then 
$1.4\ 10^8 \msun$Mpc$^{-3}$. If about half of the stars are assumed to
be in spheroids, their mass density is then
 $7\ 10^7 \msun/$Mpc$^{-3}$ and the mass density in SMBHs will be
$4\ 10^5 \msun $Mpc$^{-3}$, where the ratio M$_{\rm SMBH}$/M$_{\rm SPH}=0.006$ 
has been assumed (Magorrian et al. 1998). 
The efficiency of mass converted in FIR during a star burst 
(Firmani \& Tutukov 1994) is $5.3\ 10^{-4}$, then the FIR energy density 
produced by the star bursts of the spheroids is $4\ 10^4 \msun$Mpc$^{-3}$. 
Assuming an efficiency of 0.1 for the energy produced by the SMBH accretion, 
the energy density derived from this process is $4\ 10^4 \msun$Mpc$^{-3}$, 
an amount similar to the one derived from spheroids. The sum of both agrees 
with the total FIR energy density of $10^5 \msun$Mpc$^{-3}$ obtained by Dwek 
et al. (1998) in the DIRBE experiment.
  
A further argument supporting this scenario is given by Granato et al (2001). 
According to them, the spheroid formation rate, derived from the observed QSO 
luminosity function, leads to the observed spheroid number density at present. 
The anti-hierarchical baryonic collapse scenario of Granato et al. is consistent 
with the FIR source counts, the nature of Ly Break Galaxies (LBGs) and the 
chemical evolution of spheroids and QSOs. The SF in spheroids,  
correlated with the QSO luminosity function, shows that its bulk of activity 
happens at $z\gtrsim2$.

The previous scenario is based on some important elements. (i) Gas infall has 
to reach very high density at the radius of BH influence in order to 
produce a sufficient BH growth. (ii) The SMBH is formed by a self-regulation 
mechanism that determines its final mass and induces a natural expulsion of 
gas which reduces suddenly the optical depth. 
(iii) The bulk of SF and AGN activity is strongly enshrouded by gas and dust.

If a collision occurs between two gas-poor stellar disks the remnant 
is not sufficiently dense to be identified with a real spheroid (Hernquist 1992).
The presence of bulges in the progenitors plays an important role leading 
to a central dense spheroid comparable to the observed bulges (Hernquist 1993). 
The two SMBHs of the progenitors will coalesce spiralling-in towards the 
center by dynamical friction and gravitational wave emission, creating here 
a shallow stellar core similar to the ones observed in boxy elliptical 
galaxies (Makino \& Ebisuzaki 1996; Quinlan 1997; 
Quinlan \& Hernquist 1997; Milosavljevic et al. 2002).

\section{The Hubble sequence in the hierarchical cosmogony}

In the previous sections we have pointed out several processes which 
intervene in the evolution of luminous galaxies. Summarizing, the first
step in galaxy formation is the assembling of the dark matter halos,
which may happen either quiescently or violently, depending mostly 
on the environment. In the former case, the gas trapped inside the halos
dissipates (radiatively and turbulently), infalls to the center and forms an
inside-out disk in centrifugal equilibrium with a surface density that
depends mainly on the halo spin parameter. SF in disks is expected to be 
stationary and self-regulated by feedback and turbulent dissipation. In 
this case, the main drivers of the SF history are the gas infall and the 
angular momentum. 
Both of them are tightly related to the cosmological conditions. Some disks 
may suffer further morphological evolution due to internal gravitational 
instabilities (spiral arms, bars and bulges). 

However, the strongest changes are expected when violent major mergers occur: 
the stellar disk is heated and gas angular momentum is transferred, 
leading to the formation of spheroids and SF bursts. 
In this phase ULIGs and AGNs can appear. The products of galaxy collisions 
depend on the gas fraction, the stellar disk-to-bulge ratio and the dynamics 
of the encounter (Barnes \& Hernquist 1991, 1996; Miho \& Hernquist 1994, 1996; 
Barnes 2002). Numerical simulations show that a wide range of final products can  
be obtained, depending on these parameters. Nevertheless, the stellar component 
is almost always heated and thickened, acquiring the structure of a spheroid. 
The gas that suffers strong shocks transfers its angular momentum, and may 
originate a bulge. The gas that has not been involved in strong shocks 
retains a large amount of angular momentum and can feed an extended disk. 
Accretion from the environment supplies the spheroid/disk system with further 
angular-momentum-rich gas. 
The ultimate fate of gas is to be converted into stars. 
When the gas surface density is sufficiently high, the disk stellar population is 
practically coeval to the bulge, but as the gas surface density is lower, the
slower is the SF rate. 
 
In the hierarchical framework, the major merging rate is easily estimated by the 
extended Press-Schechter formalism or by N-body simulations (e.g., Gottloeber 
et al. 2001) and, for a given mass, is increasing with z. 
Multiplying the merging rate by the halo mass function at different $z$'s, one 
obtains that the merging rate density (per unit of volume) for halos with masses 
smaller than $10^{13}\msun$ has its maximum at $z>1$; for halos in clusters, 
the maximum is shifted to earlier epochs. Therefore, most of spheroids, upon 
the understanding that they arise from major mergers, are born at high redshifts, 
when galaxies are still gas rich. The smaller fraction 
of major mergers at late epochs occurs mainly between gas-poor galaxies with an 
important spheroidal component. Around the spheroid a disk grows by gas 
accretion according to the conditions of the environment and by the energy 
feedback of the spheroid, the bulge-to-disk mass ratio being an indicator of 
the Hubble type. When the spheroid is more massive than the disk, we have 
an elliptical or S0 galaxy. The lack of a massive disk in these cases may be 
due to the facts that {\it (i)} the merger has occurred recently, {\it (ii)} 
the gas accretion has been inhibited, or {\it (iii)} the environment is poor
in gas. The existence of field elliptical galaxies with an old stellar 
population may be indicative of conditions that inhibit the formation and 
growth of a disk. 

Using semi-analytic modelation and taking into account many of the physical
processes mentioned above, Cole et al. (2000) have made predictions
about the morphology mixing of galaxies and the dependence on
the environment. Their models reproduce the relative number of E+S0 
galaxies. However, concerning the age of spheroids, the situation is 
not so clear. Their predicted colour-magnitude relation for ellipticals in
clusters is significantly flatter than that observed at bright magnitudes. 
Benson et al. (2002) have compared the same models with samples of field 
spheroidal galaxies with redshifts up to two. Models reveal some difficulty 
predicting the observed red population of E+S0 galaxies, while on the blue 
side the agreement is satisfactory. Colours reflect only 
approximately the ages of old spheroids. The ages of
elliptical galaxies derived by different observational
methods, such as those mentioned in section 5.1, are difficult to
reconcile with the model predictions. Again, the situation could be
alleviated  if gas accretion and disk growth around
spheroids would be inhibited (item (ii) above), in this way allowing some 
of the spheroids to evolve with a negligible SF activity. 
A careful comparison between models and observations
of the relation between the intrinsic colours of the
disk and the Hubble type should be of some relevance.
In fact, a growing disk around a spheroid would evolve
inside-out from blue to red while the bulge-to-disk
ratio decreases; however, this situation depends on the
physical conditions created by the collision as well
by gas accretion and the intense feedback.

Back to galaxy collisions, it should be noted that its stochastic 
nature (as well as the extended redshift range of spheroid formation) contribute 
to the scatter of the FJR and KR, and possibly of the FP. 
While the scatter of TFR is a faithful product of the stochastical properties 
of the primordial density fluctuations (\S 4.1), the scatters of the spheroid 
scaling relations (FJR, KR, FP) are further increased by the astrophysical 
processes in which they have been involved (collisions, feedback).

\subsection{Connecting spheroid formation with QSOs}

The scenario described above opens the possibility to understand the physics of 
the HS, its dependence on the environment, the nature of peculiar elliptical 
galaxies with extended low surface density gaseous disks, as well as the 
connection with high redshift objects like QSOs and submillimeter sources. 

QSOs provide now an important probe up to $z \simeq 6$. Unfortunately, their 
lifetime and their connection with the host galaxy are still rather uncertain.
An approach to the bulge-AGN evolution, in the context of the hierarchical
scenario, has been proposed by e.g.,  Cattaneo et al. (1998), Cattaneo 
(1999, 2002), Kauffmann \& Haehnelt (2000), and Wyithe $\&$ Loeb (2002). 
In these semi-analytical models, several assumptions and prescriptions are used 
to calculate the gas fractions in the merging disks (related to gas infall
 and SF) and the lifetime of QSOs. 
The models are able to reproduce roughly the ratio $M_{\rm SMBH}/M_{\rm bulge}$ 
and the evolution of the QSO luminosity function. The major difficulty arises 
at the low redshift side: the observed QSO population decreases dramatically 
since $z\sim 2$ to $z=0$, while the merging rate decreases only a little. 
Cavaliere $\&$ Vittorini (2000) have studied this problem. An important fact to 
take into account is that the gas fraction decreases in galaxies with time. 
Furthermore, since SF in the inner disk regions is faster than at the periphery, 
the low angular momentum gas ---which feeds the SMBH--- is exhausted at a rate 
higher than the average. Using complete galaxy evolutionary models (e.g., 
Firmani \& Avila-Reese 2000), this effect could be included into the QSO 
models to explain the luminosity evolution of QSOs at low $z$'s. 
Another interesting aspect has to do with the multipliciy of processes able to 
activate SMBH. While it is reasonable to imagine that at early times QSOs have 
been activated mainly by major mergers, later on other processes besides 
mergers may become competitive in the inner region of a galaxy to stimulate a 
gas inflow and a SMBH feeding. 
The environment seems to be a key ingredient. Observations show that QSOs were 
more clustered at high redshifts than at low ones.

\section{Concluding remarks}

The study of galaxies is not anymore only a ``taxonomical'' task. We are
arriving now at a ``genetical'' level of their understanding, which opens 
the possibility to explain their diversity, in particular the origin
of the HS. A relevant question emerges: is the origin of the HS dominated
by cosmological conditions (nature) or by ulterior astrophysical 
processes (nurture)?  The inductive approach has
been particularly useful for exploring the latter item. A key aspect
is the identification and the interconnection of the astrophysical 
processes that drive changes in the Hubble type, as well as the 
identification of observable fossil records of the different evolutionary phases.
Models of spectrophotometric, chemical and dynamical evolution 
are the instruments which helped to reconstruct the past history from the 
observed fossil records. Most of the results from studies related to the
inductive approach showed the necessity to connect galaxy evolution to 
external conditions. These conditions can be given by a cosmological 
scenario of structure formation. 

The success of the inflationary CDM cosmology has pushed researchers to 
explore their ultimate consequences regarding
galaxy formation (deductive approach). The predictions of models
and simulations are encouraging; the hierarchical CDM model offers an 
invaluable theoretical background for understanding galaxies. The formation and 
evolution of disk galaxies within the CDM halos   (\S 2)
has been modeled with a minimum of free astrophysical parameters (\S 4).
Therefore, the confrontation between theory and observations is 
particularly promising in this case. It seems that most of the properties and
correlations of disk galaxies, including the correlations along the HS, 
are strongly linked to the initial conditions of the hierarchical CDM
cosmology. A potential problem of the inside-out disks is their rapid
size evolution, which could not be confirmed by high-redshift observations.
 More detailed comparisons of models and observations are needed at 
all the levels. They will allow to trace some properties of the early 
universe, as well as to infer the nature of the mysterious dark matter. 
As a result of some of these comparisons, galaxy-sized CDM halos 
seem to be too cuspy and with too much substructure. If the dark particles 
are assumed with some self-interacting properties or warm instead of cold, 
then the agreement with observations improves (\S 2). 

The situation concerning spheroids is more complex than for disks. 
The age of the isolated spheroids and the details of the disk growing 
around the bulges are among the issues which keep open the debate 
about the congruency of the hierarchical cosmogony regarding
spheroid formation.
While the backbone of spheroid formation might be again the dark 
matter processes (major mergers mainly), complex astrophysical processes 
(e.g., non-stationary SF, strong feedback and cooling, dynamical 
processes, angular momentum transfer, SMBHs, gas outflow-inflow phenomena, 
etc.) make difficult the definition of an unambiguous evolutionary sequence. 
For spheroids, the astrophysical processes seem to be dominant. 
As has been emphasized by Renzini (1994), to reveal the complexity of 
galaxy evolution the deductive approach has to go necessarily together 
with the inductive one.  

Great perspectives are open for further research in galaxy formation and 
evolution. A better understanding of all kinds of astrophysical processes
involved in the evolution of galaxies is crucial. Star formation, 
hydrodynamics and feedback are among the most relevant, in particular, 
for spheroids. A better modelling of the luminous objects which form 
within the dark halos, will allow us to use the observational properties of 
galaxies to trace the physical conditions of the early universe and dark matter. 
From the astrophysical side, the connection of spheroid formation to ULIGs, 
submillimeter sources and QSO-AGN phenomenon is an appealing problem. 
The interplay of galaxy evolution with the IGM is also an important avenue 
of research.  

\vspace{0.5cm}

As Ivan King said sometime ago, {\it the main challenge is reserved for 
those who obtain results linking theory and observations; they are the 
astronomers.}

\acknowledgments

This work was supported by CONACyT grant 33776-E to V.A.

\end{document}